\documentclass[apj]{emulateapj} 

\usepackage{epsfig}
\usepackage{amsmath}
\usepackage{amssymb}
\usepackage{natbib}
\usepackage{threeparttable} 
\usepackage{booktabs}

\usepackage{epstopdf}

\newcommand{\chan}{\textit{Chandra}}
\newcommand{\swift}{\textit{Swift}}

\newcommand{\xmm}{\textit{XMM-Newton}}
\newcommand{\inte}{\textit{Integral}}

\newcommand{\beppo}{\textit{BeppoSAX}}

\newcommand{\Msun}{\mathrm{M}_{\odot}}
\newcommand{\lum}{\mathrm{erg~s}^{-1}}
\newcommand{\flux}{\mathrm{erg~cm}^{-2}~\mathrm{s}^{-1}}

\newcommand{\cnts}{\mathrm{counts~s}^{-1}}
\newcommand{\mdot}{\mathrm{M_{\odot}~yr}^{-1}}
\newcommand{\nh}{\mathrm{cm}^{-2}}

\newcommand{\source}{J2224}
\newcommand{\sourcefull}{SAX~J2224.9+5421}
\newcommand{\exo}{EXO 1745--248}
\newcommand{\sax}{SAX J1808.4--3658}

\newcommand{\swiftpulsar}{Swift J1749.4--2807}

\newcommand{\saxeen}{SAXJ1828.5--1037}
\newcommand{\saxtwee}{SAXJ1753.5--2349}
\newcommand{\saxdrie}{SAXJ1806.5--2215}

\newcommand{\radiopulsar}{IGR J18245--2452}

\newcommand{\atel}{ATel}
\slugcomment{Accepted to ApJ}

\shorttitle{\sourcefull\ in quiescence}
\shortauthors{Degenaar, Wijnands \& Miller}

\begin{document}

\title{The Quiescent Counterpart of the Peculiar X-Ray Burster \sourcefull}

\author{N. Degenaar$^{1,}$\altaffilmark{3}, R. Wijnands$^{2}$, J. M. Miller$^{1}$
}
\affil{$^1$Department of Astronomy, University of Michigan, 500 Church Street, Ann Arbor, MI 48109, USA; degenaar@umich.edu\\
$^2$Astronomical Institute Anton Pannekoek, University of Amsterdam, Postbus 94249, 1090 GE Amsterdam, The Netherlands
}
\altaffiltext{3}{Hubble Fellow}

\begin{abstract}
\sourcefull\ is an extraordinary neutron star low-mass X-ray binary. Albeit discovered when it exhibited a $\simeq$10-s long thermonuclear X-ray burst, it had faded to a 0.5--10 keV luminosity of $L_{\mathrm{X}}\lesssim8\times10^{32}~(D/\mathrm{7.1~kpc})^2~\lum$ only $\simeq$8~hr later. It is generally assumed that neutron stars are quiescent (i.e., not accreting) at such an intensity, raising questions about the trigger conditions of the X-ray burst and the origin of the faint persistent emission. We report on a $\simeq$51 ks \xmm\ observation aimed to find clues explaining the unusual behavior of \sourcefull. We identify a likely counterpart that is detected at $L_{\mathrm{X}}$$\simeq$$5\times10^{31}~(D/\mathrm{7.1~kpc})^2~\lum$ (0.5--10 keV) and has a soft X-ray spectrum that can be described by a neutron star atmosphere model with a temperature of $kT^{\infty}\simeq50$~eV. This would suggest that \sourcefull\ is a transient source that was in quiescence during our \xmm\ observation and experienced a very faint (ceasing) accretion outburst at the time of the X-ray burst detection. We consider one other potential counterpart that is detected at $L_{\mathrm{X}}$$\simeq$$5\times10^{32}~(D/\mathrm{7.1~kpc})^2~\lum$ and displays an X-ray spectrum that is best described by power law with a photon index of $\Gamma \simeq 1.7$. Similarly hard X-ray spectra are seen for a few quiescent neutron stars and may be indicative of a relatively strong magnetic field or the occurrence of low-level accretion.
\end{abstract}


\section{Introduction}
When matter accretes onto the surface of a neutron star, it can undergo unstable thermonuclear burning resulting in a brief, intense flash of X-ray emission. Such thermonuclear X-ray bursts (type-I X-ray bursts; shortly X-ray bursts hereafter) are observed from low-mass X-ray binaries (LMXBs), in which a neutron star accretes matter from a sub-solar companion star that overflows its Roche lobe. X-ray bursts are a potential tool to constrain the fundamental properties of neutron stars \citep[e.g.,][]{vanparadijs1979,steiner2010,suleimanov2011_eos}, and can give valuable insight into the accretion flow around the compact object \citep[e.g.,][]{yu1999,ballantyne2004,zand2012,degenaar2013_igrj1706,worpel2013}.

Wide-field monitoring with observatories such as \beppo, \inte, and \swift\ has led to the discovery of X-ray bursts for which no persistent emission could be detected above the instrument background \citep[e.g.,][]{zand1999_burstonly,cocchi2001,cornelisse02,chelovekov07_ascabron,delsanto07,wijnands09,linares09,degenaar2010_burst,degenaar2011_burst, degenaar2013_igrj1706}. This implies that these \textit{burst-only sources} are accreting at 2--10 keV luminosities of $L_{\mathrm{X}}\lesssim10^{36}~\lum$. This is lower than typically seen for neutron star LMXBs \citep[e.g.,][]{chen97,wijnands06,campana09,degenaar2012_gc}. The burst-only sources trace a relatively unexplored accretion regime and  provide valuable new input for thermonuclear burning models \citep[e.g.,][]{cooper07,peng2007,degenaar2010_burst}. 

\begin{figure*}
 \begin{center}
\includegraphics[width=8.0cm]{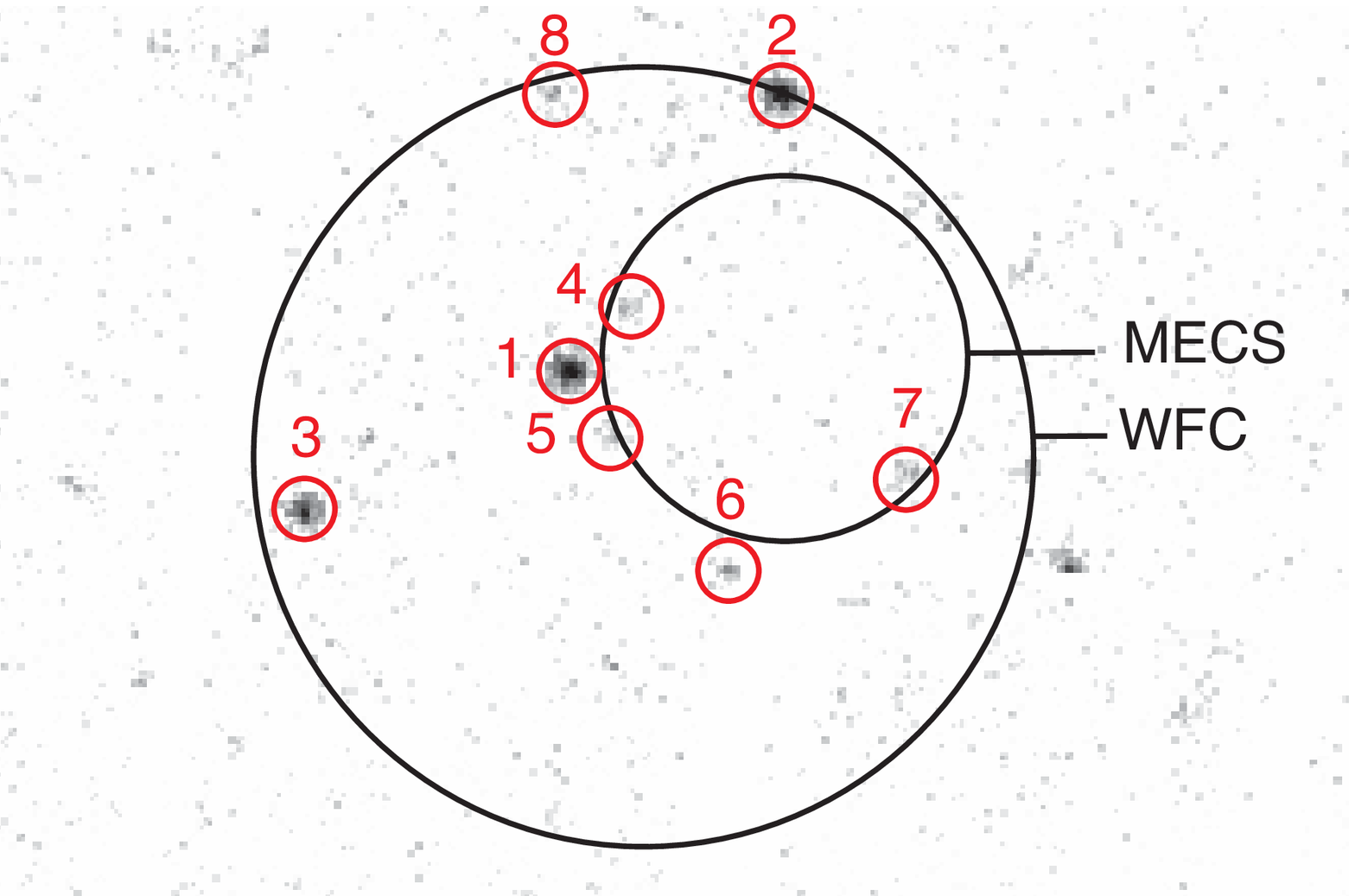}\hspace{0.5cm}
\includegraphics[width=8.0cm]{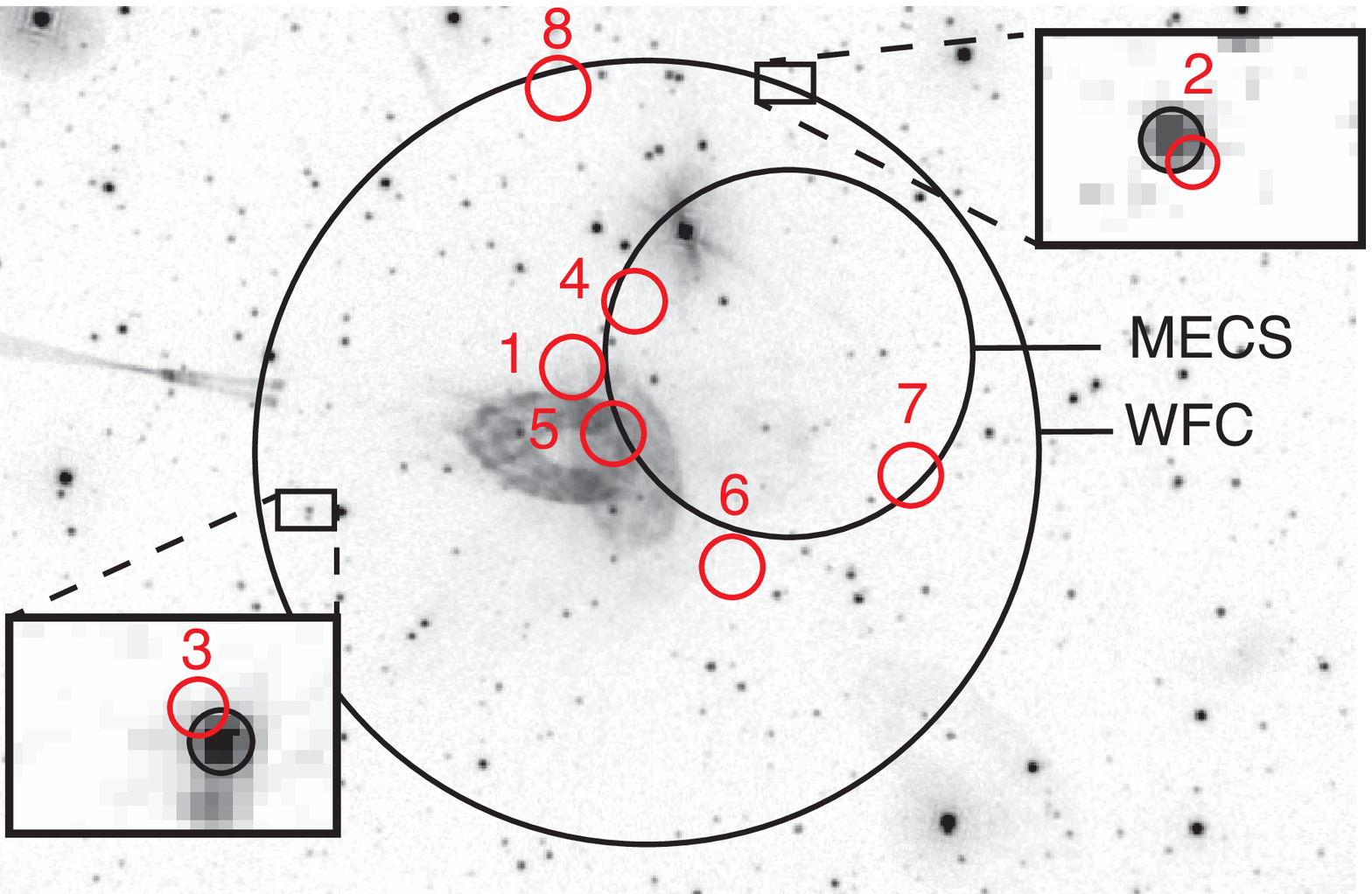}
    \end{center}
\caption[]{{\xmm\ images of the field around \source. Left: Combined EPIC X-ray image (0.5--10 keV). The \beppo/WFC positional uncertainty of the X-ray burst ($3.2'$) and that of the tentative counterpart identified in follow-up \beppo/MECS observations ($1.5'$) are indicated. Within the WFC error circle there were eight faint X-ray sources detected with all three EPIC detectors, which are indicated by numbered red circles ($15''$ in size). Right: OM $V$-band image. The insets show magnifications to indicate possible associations with optical objects (black circles). In those sub-images the size of the red circles correspond to the X-ray positional uncertainties of source 2 and 3 ($1.7''$ and $1.6''$, respectively). The ellipsoidal feature in the center of the image concerns an artifact (straylight). }}
 \label{fig:image}
\end{figure*}

Follow-up observations with more sensitive narrow-field instruments (e.g., onboard \beppo, \chan, \xmm\ and \swift) 
have revealed that the burst-only sources fall into two categories: 1) LMXBs accreting (quasi-) persistently at $L_{\mathrm{X}}\simeq10^{34-35}~\lum$ \citep[e.g.,][]{zand05,chelovekov07_ascabron,delsanto07,degenaar2010_burst}, and 2) transient LMXBs that exhibit weeks to months long outbursts of $L_{\mathrm{X}}\simeq10^{34-36}~\lum$, but are otherwise quiescent \citep[e.g.,][]{cocchi2001,cornelisse02,hands04,wijnands09,campana09}. 

In quiescence, neutron star LMXBs are dim with $L_{\mathrm{X}}\simeq10^{31-33}~\lum$ \citep[e.g.,][]{campana1998,jonker2004,heinke2009,guillot2013}. The quiescent spectra typically contain a soft component that is well-fitted by a neutron star atmosphere model. This emission is ascribed to thermal radiation from the surface of the neutron star, and may act as a probe of its interior properties \citep[e.g.,][]{vanparadijs1984,brown1998,rutledge1999,wijnands2004,cackett2006,degenaar2013_ter5}. With their low mass-accretion rates, the burst-only sources can provide valuable new insight into the thermal evolution of neutron stars \citep[][]{wijnands2012}. 

An additional, non-thermal spectral component is often present in the quiescent spectra. It can be described by a simple power law with a photon index of $\Gamma\simeq1-2$. This component has been associated with (stochastic) intensity variations and therefore tentatively ascribed to the presence of a residual accretion flow \citep[e.g.,][]{rutledge2002_aqlX1,cackett2005,cackett2011_aqlx1,fridriksson2011,degenaar2012_1745,bernardini2013,wijnands2013}. Since there are some indications that the hard power-law emission is particularly prominent in the quiescent spectra of accreting millisecond X-ray pulsars \citep[AMXPs; e.g.,][]{wijnands05_amxps,campana2008,heinke2009,degenaar2012_amxp,linares2013_M28}, it has also been explained as the result of accretion onto the magnetosphere of the neutron star, or a shock from a pulsar wind colliding with matter flowing out of the donor star \citep[e.g.,][]{campana1998,rutledge2001,linares2013_M28}.

\subsection{The Peculiar X-Ray Burster \sourcefull}
Perhaps the most tantalizing burst-only source is \sourcefull\ (\source\ hereafter). It was detected with the \beppo\ Wide Field Camera (WFC) as a $\simeq$10-s long X-ray flash on 1999 November 6 \citep[][]{gandolfi1999}. Initially dubbed GRB 991106, multi-wavelength follow-up observations failed to detect a fading afterglow and cast doubt on a GRB nature \citep[e.g.,][]{antonelli1999,gandolfi1999_3,frail1999,jensen1999,gorosabel1999}. \citet{cornelisse02} demonstrated that the properties of the event were consistent with a thermonuclear X-ray burst, which would identify \source\ as a new neutron star LMXB. A source distance of $D \lesssim 7.1$~kpc was inferred by assuming that the X-ray burst peak did not exceed the Eddington limit ($L_{\mathrm{edd}}=2\times10^{38}~\lum$).

No accretion emission could be detected with the WFC around the time of the X-ray burst, implying a 2--28 keV luminosity of $L_{\mathrm{X}}\lesssim2\times10^{36}~(D/\mathrm{7.1~kpc})^2~\lum$ \citep[][]{cornelisse02}. Rapid follow-up observations with the \beppo\ Medium-Energy Concentrator Spectrometer (MECS), performed $\simeq$8~hr after the X-ray burst detection, revealed only one source within the WFC error circle that was detected at a 0.5--10 keV luminosity of $L_{\mathrm{X}}\simeq8\times10^{32}~(D/\mathrm{7.1~kpc})^2~\lum$ \citep[][]{antonelli1999}. This raises the question whether the X-ray burst was ignited when the neutron star was accreting at a very low level, or whether it exhibited a faint, undetected accretion outburst that had ceased within 8~hr of the X-ray burst detection. 

In this work we present a deep \xmm\ observation to search for an X-ray counterpart of \source, and to find clues to its nature and unusual behavior.

\begin{table*}
\begin{center}
\caption{Positions and Basic Properties of Detected X-Ray Sources.\label{tab:sources}}
\begin{tabular*}{0.99\textwidth}{@{\extracolsep{\fill}}lcccccc}
\hline
Source &  R. A. & Dec. & Error & MOS Count Rate  & PN Count Rate & PN Hardness Ratio \\
 &  (J2000) & (J2000) & $('')$ & $(10^{-3}~\cnts)$ & $(10^{-2}~\cnts)$ &  \\
\hline
1 \dotfill & 22 24 52.94 & +54 22 38.2 & 1.6 & $3.9 \pm 0.3$ & $1.3 \pm 0.1$ & $0.94 \pm 0.17$   \\
2 \dotfill & 22 25 07.87 & +54 21 30.2 & 1.7 & $3.2 \pm 0.3$ & $0.8 \pm 0.1$ & $0.73 \pm 0.15$    \\
3 \dotfill & 22 24 40.99 & +54 24 54.4 & 1.6 & $3.8 \pm 0.3$ & $1.5 \pm 0.1$ & $0.06^{+0.37}_{-0.06}$    \\
4 \dotfill & 22 24 49.65 & +54 23 10.1 & 2.3 & $0.7\pm0.2$ & $0.17 \pm 0.03$ &  $0.22^{+0.58}_{-0.22}$   \\
5 \dotfill & 22 24 50.64 & +54 22 05.1 & 2.4 & $0.6\pm0.2$  & $0.13 \pm 0.03$ & $1.22 \pm 0.46$    \\
6 \dotfill & 22 24 43.99 & +54 20 59.9 & 2.2 & $0.6\pm0.2$  & $0.23 \pm 0.05$ & $1.54 \pm 0.52$    \\
7 \dotfill & 22 24 33.82 & +54 21 48.9 & 2.2 & $0.7\pm0.2$  & $0.24 \pm 0.04$ & $1.99 \pm 0.34$    \\
8 \dotfill & 22 24 53.74 & +54 24 54.9 & 2.1 & $0.6\pm0.2$  & $0.23 \pm 0.04$ & $2.29 \pm 0.32$    \\
\hline
\end{tabular*}
\tablecomments{The quoted positional uncertainties are at 90\% confidence level, and are a combination of the statistical error from the detection algorithm and an estimated systematic uncertainty of $1.5''$ \citep[][]{watson2009}. Count rates errors are at the 1$\sigma$ level of confidence. We consider sources 1 and 4 as possible counterparts for \source.
}
\end{center}
\end{table*}

\section{Observations and Analysis}

\subsection{X-Ray Data}
The field around \source\ was observed with \xmm\ for $\simeq$51~ks on 2013 May 29 UT 06:04--20:24 (ID 0720880101). X-ray data was obtained with the European Photon Imaging Camera (EPIC), which consists of two MOS detectors \citep[][]{turner2001_mos}, and one PN camera \citep[][]{struder2001_pn}. All instruments were operated in the full frame imaging mode with the medium optical blocking filter applied. We reduced and analyzed the data using the Science Analysis Software (\textsc{sas}; version 11.0). The Original Data Files (ODF) were reprocessed using the tasks \textsc{emproc} and \textsc{epproc}. The observation was free from background flaring. 

Figure~\ref{fig:image} (left) displays the composite X-ray image of all three detectors. There are eight faint X-ray sources within the \beppo/WFC error circle that were detected with all three EPIC instruments. The source positions (determined using the task \textsc{edetect$\_$chain}) are listed in Table~\ref{tab:sources}. Sources 1 and 3 were previously detected at similar intensities in \swift/XRT data of the field \ \citep[named SAX J2225-1 and SAX J2225-2, respectively;][]{campana09}. 

We determined the count rate for each source by employing the task \textsc{eregionanalyse}, using circular regions with radii of $20''$ (400 pixels) centered on the source positions. For the background we used a circular region with a radius of $60''$ placed on a source-free part of the CCD. Source and background spectra were extracted using \textsc{especget}, which also generated the response files. We co-added the two MOS spectra and their response files with \textsc{epicspeccombine}. Using \textsc{grppha}, the spectral data was grouped to contain a minimum of 15 photons per bin. We used \textsc{XSpec} \citep[version 12.7;][]{xspec} to perform spectral fits in an energy range of 0.5--10 keV.

\subsection{X-Ray Spectral Analysis}
To obtain a rough characterization of the spectral shape, we determined the hardness ratio for each source using the PN data. For the present purpose, we defined this quantity as the ratio of counts in the 2--10 and 0.5--2 keV bands. This information is included in Table~\ref{tab:sources}. 

We explored their X-ray spectra using two models that are widely used for neutron star LMXBs at low luminosities; a simple power law (\textsc{pegpwrlw} in \textsc{XSpec}), and a neutron star atmosphere model \citep[for which we chose \textsc{nsatmos};][]{heinke2006}. For the latter we always fixed the neutron star mass ($M=1.4~\Msun$) and radius ($R=10$ km), as well as the distance ($D=7.1$ kpc) and the normalization (unity). The neutron star temperature was then the only free fit parameter. Thermal bolometric fluxes were estimated by extrapolating the \textsc{nsatmos} model fits to the 0.01--100 keV energy range. In all spectral fits we accounted for interstellar absorption by including the \textsc{tbabs} model in \textsc{XSpec}, using the \textsc{wilm} abundances and \textsc{vern} cross-sections \citep[][]{verner1996,wilms2000}.

\subsection{Optical and Ultra-Violet Data}
Quasi-simultaneous optical and ultra-violet (UV) coverage was provided by the Optical Monitor \citep[OM;][]{mason1996}. The position of \source\ was observed with the OM operated in image mode. Exposures of 2000, 2460, and 2200~s were obtained using the $V$ ($\lambda_{\mathrm{c}}\simeq5407$~\AA), $UVW1$ ($\lambda_{\mathrm{c}}\simeq2905$~\AA), and $UVW2$ ($\lambda_{\mathrm{c}}\simeq2070$~\AA) filters, respectively. Reduction, source detection and photometry was performed using the meta-task \textsc{omichain}. 

None of the X-ray sources within the \beppo\ uncertainty of \source\ are detected in the $UVW2$ image. However, source 2 and 3 both have a potential counterpart in the $V$ and $UVW1$ wavebands. This is illustrated by Figure~\ref{fig:image} (right), which shows the OM $V$-band image.\\

\begin{figure*}
 \begin{center}
\includegraphics[width=8.7cm]{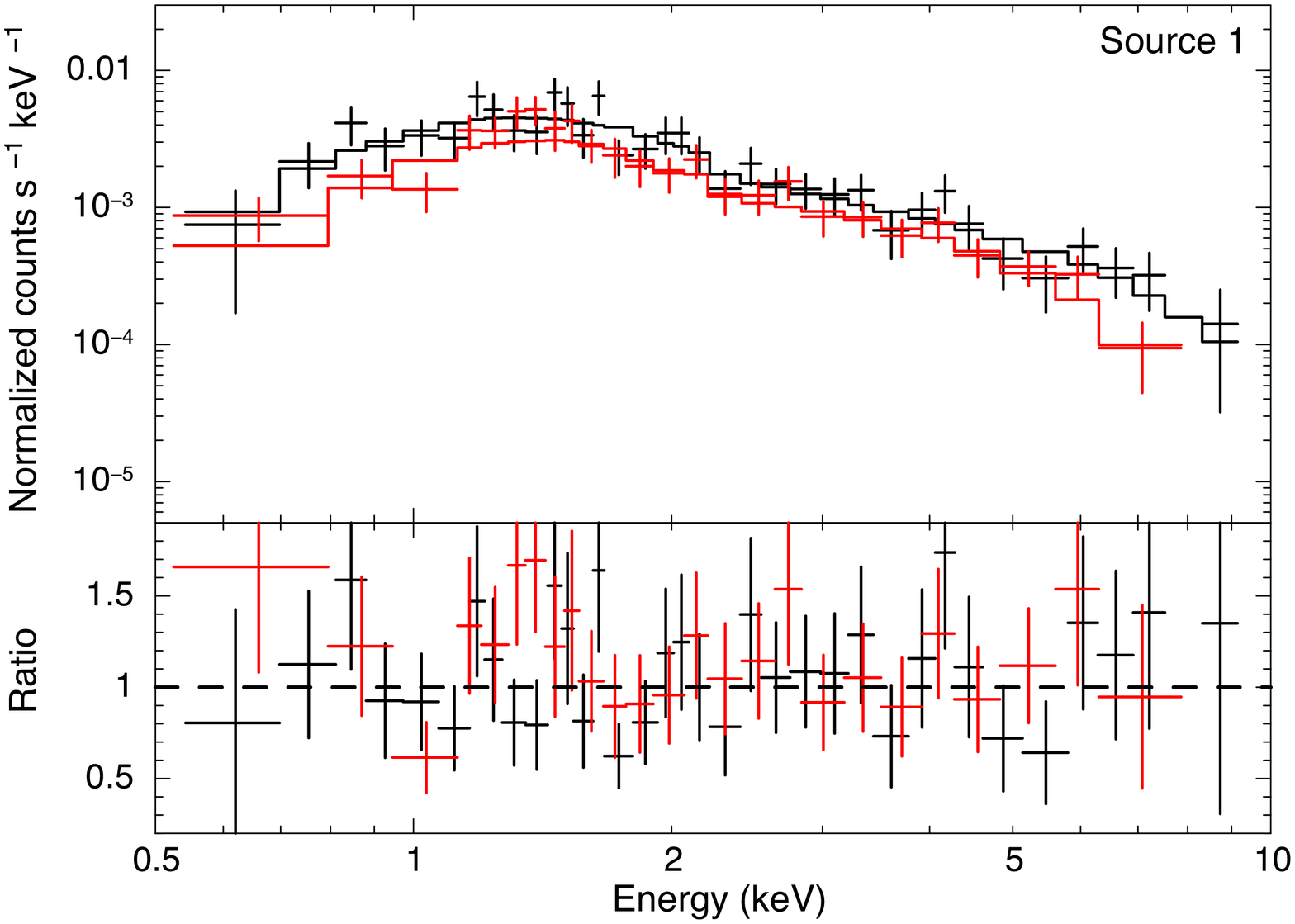}\hspace{+0.5cm}
\includegraphics[width=8.7cm]{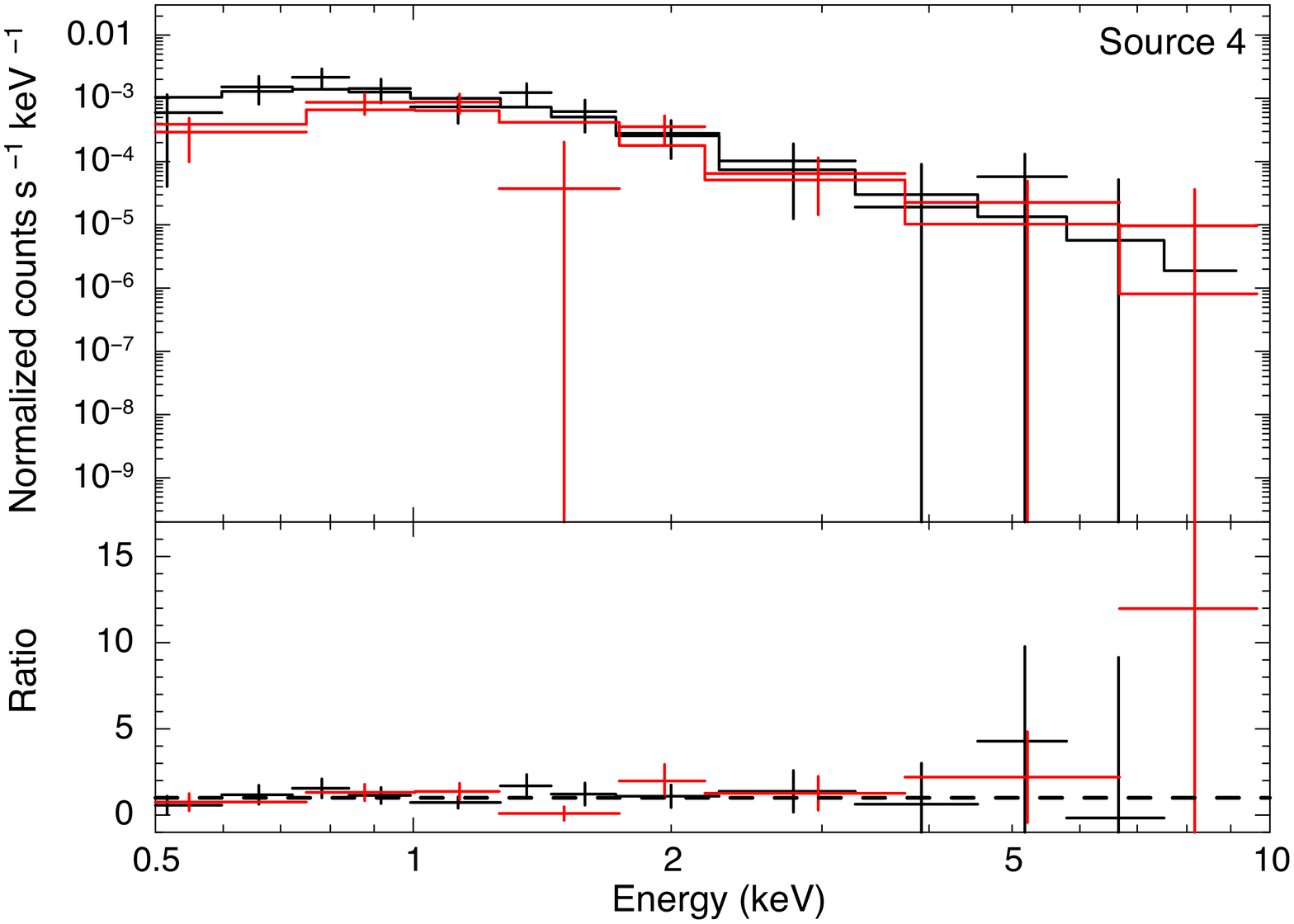}
    \end{center}
\caption[]{{\xmm\ PN (black) and combined MOS1/MOS2 (red) spectra of the two candidate counterparts of \source. The solid lines indicate the model fits; a power law for source 1 and a neutron star atmosphere model for source 4 (see Section~\ref{sec:results}). The bottom panels show the model to data ratio.}}
 \label{fig:spec}
\end{figure*}

\section{Results}\label{sec:results}

\subsection{Sources 5--8: Too X-Ray Hard}\label{subsec:srcother}
Sources 5--8 have the hardest spectra among the detected sources, as is indicated by their high hardness ratio's (Table~\ref{tab:sources}). Although a limited number of counts prohibits a detailed analysis, their spectra can be described by a power law with photon indices of $\Gamma\simeq1-3$ and hydrogen column densities that are well in excess of the Galactic value in the direction of \source\ \citep[$N_H \simeq 4.6\times10^{21}~\nh$;][]{kalberla2005}. This renders them unlikely counterparts to the X-ray burster. Their unabsorbed fluxes lie in a range of $F_{\mathrm{X,unabs}} \simeq (1-10)\times10^{-14}~\flux$ (0.5--10 keV). 

It is more likely that these hard X-ray sources are (obscured) background Active Galactic Nuclei (AGN) or Galactic accreting white dwarfs (i.e., cataclysmic variables, CVs; see also Section~\ref{subsec:srctwothree}). Both types of objects are abundant in the sky. For example, assuming a number density of $\simeq3\times10^{-5}$~pc$^{-3}$ \citep[e.g.,][]{schwope2002}, we expect $\simeq3$ CVs within the WFC error circle. Furthermore, down to a flux of $\simeq10^{-14}~\flux$, about $3\times10^{2}$ AGN are expected per square degree of sky \citep[][]{brandt2005}. That implies that within the WFC uncertainty, roughly 3 AGN with fluxes of $\gtrsim10^{-14}~\flux$ are expected.

\subsection{Sources 2 and 3: Too Optically/UV Bright}\label{subsec:srctwothree}
Sources 2 and 3 are the only ones with possible optical/UV associations. Based on the total number of $V$-band objects detected within the $17'\times17'$ OM field-of-view, we estimate the probability of a chance alignment within the $1.6''$ and $1.7''$ positional uncertainties of source 2 and 3 to be $\simeq5\times10^{-3}$ and $\simeq6\times10^{-3}$, respectively. 

For source 2 we measure AB-magnitudes of $V=16.41\pm0.01$~mag and $UVW1=17.45\pm0.05$~mag. The corresponding flux densities are $f_{\mathrm{V}} = (1.1\pm0.1) \times 10^{-15}~\flux$~\AA$^{-1}$ and $f_{\mathrm{UVW1}} =  (3.8\pm0.2) \times 10^{-16}~\flux$~\AA$^{-1}$ (uncorrected for reddening). The X-ray spectrum of source 2 cannot be described by a neutron star atmosphere model ($\chi_{\nu}^2 > 3$), but fits to a power law with $\Gamma= 2.0 \pm 0.3$ and $N_H \simeq (6.9\pm2.3)\times10^{21}~\nh$. The resulting 0.5--10 keV unabsorbed flux is $F_{\mathrm{X,unabs}} \simeq (4.9 \pm 0.7) \times10^{-14}~\flux$. 

By multiplying the obtained $V$ flux density with the width of the filter ($\simeq$684~\AA), we obtain a ratio of the X-ray to optical flux of $F_{\mathrm{X}}/F_{\mathrm{V}}\simeq$0.3 for source 2. Quiescent neutron star LMXBs typically have a much higher X-ray flux compared to that in the optical/UV band \citep[a ratio of $\simeq$10 or higher, e.g.,][]{homer2001,russell2006,hynes2012}. We therefore consider it unlikely that source 2 is the counterpart to \source. It is plausible that it is a background AGN (see also Section~\ref{subsec:srcother}). The X-ray spectra of these objects can typically be described by a $\Gamma \simeq 2$ power-law model, and their X-ray/optical flux ratio's fall in a wide range of $F_{\mathrm{X}}/F_{\mathrm{V}}\simeq0.1-10$.

Source 3, for which we measure $V=18.08\pm0.07$~mag ($f_{\mathrm{V}} =  (2.3\pm0.2) \times 10^{-16}~\flux$~\AA$^{-1}$) and $UVW1=19.10\pm0.19$~mag ($f_{\mathrm{UVW1}} = (8.4\pm1.4) \times 10^{-17}~\flux$~\AA$^{-1}$), also seems too optically/UV bright for a quiescent neutron star LMXB. This is supported by the fact that neither a power law nor a neutron star atmosphere model, or a combination of the two, can satisfactory describe the X-ray spectrum of source 3 (all trials resulted in $\chi_{\nu}^2 > 1.5$ for 53--55 d.o.f.). The data is instead better fitted with e.g., a combination of a black body and an optically thin thermal plasma model (\textsc{mekal}; yielding $\chi_{\nu}^2 = 1.07$ for 52 d.o.f.), or a double \textsc{mekal} model ($\chi_{\nu}^2 = 1.14$ for 52 d.o.f.). Such models are often used to describe the spectra of (magnetic) CVs \citep[e.g.,][]{ramsay2004,baskill2005}. Indeed, we expect several CVs present within the WFC error circle of \source\ (see Section~\ref{subsec:srcother}).

\subsection{Source 4: An X-Ray Soft Candidate Counterpart}\label{subsec:srcfour}
Source 4 stands out by showing a relatively soft X-ray spectrum that can be described by an absorbed power-law model with $\Gamma=2.9^{+2.2}_{-1.0}$ and $N_H = 2.8^{+4.5}_{-2.2}\times10^{21}~\nh$ (Table~\ref{tab:spec}). The obtained hydrogen column density is consistent with the Galactic value in the direction of \source. A neutron star atmosphere model provides a good fit, yielding a neutron star temperature (as seen by an observer at infinity) of $kT^{\infty} = 50\pm 4$~eV and $N_H = (4.8\pm0.2)\times10^{21}~\nh$. The obtained 0.5--10 keV luminosity is $L_{\mathrm{X}}\simeq 5\times10^{31}~(D/\mathrm{7.1~kpc})^2~\lum$. The X-ray spectrum of source 4 is shown in Figure~\ref{fig:spec} (right), and the fit results are summarized in Table~\ref{tab:spec}.

The spectral shape of source 4 is typical for a quiescent neutron star LMXB. Combined with its location near the center of the \beppo/WFC error circle, and lack of an optical/UV association makes it a strong candidate counterpart to \source. The intensity inferred from our \xmm\ observation is well below the sensitivity of previous X-ray observations that covered the source region \citep[e.g., with \beppo\ and \swift;][]{antonelli1999,campana09}, which explains why this object was not found before.

\subsection{Source 1: An X-Ray Hard Candidate Counterpart}\label{subsec:srcone}
A neutron star atmosphere model fails to describe the spectral data of source 1 ($\chi_{\nu}^2 > 3$). However, a power law provides an adequate fit, yielding $\Gamma= 1.7 \pm 0.2$ and $N_H \simeq (6.7\pm1.9)\times10^{21}~\nh$ (comparable to the Galactic value toward \source). The inferred 0.5--10 keV luminosity is $L_{\mathrm{X}}\simeq 5\times10^{32}~(D/\mathrm{7.1~kpc})^2~\lum$. The X-ray spectrum of source 1 is shown in Figure~\ref{fig:spec} (left).

Although the spectral shape of source 1 is harder than that typically observed for quiescent neutron star LMXBs, there are a few sources with similar properties (see Section~\ref{sec:discussion}). Combined with its location near the center of the \beppo/WFC error circle, and lack of a bright optical/UV association, we cannot discard source 1 as a potential counterpart. This object was detected during \swift/XRT observations obtained in 2006--2008 at $L_{\mathrm{X}}\simeq8\times10^{32}~(D/\mathrm{7.1~kpc})^2~\lum$ \citep[][]{campana09}, i.e., similar  as during our \xmm\ observation.

We probed the upper limits on any thermal emission by adding an \textsc{nsatmos} component to the best power-law fit. This yielded $kT^{\infty} \lesssim 61$, and a fractional contribution of the thermal component to the total unabsorbed 0.5--10 keV model flux of $\lesssim$24\%. The spectral results of source 1 are listed in Table~\ref{tab:spec}.

\begin{table*}
\begin{center}
\caption{Spectral Results of the Two Candidate X-Ray Counterparts.\label{tab:spec}}
\begin{tabular*}{0.99\textwidth}{@{\extracolsep{\fill}}lcccc}
\hline
 & \multicolumn{2}{c}{Source 1}  & \multicolumn{2}{c}{Source 4} \\
Parameter (unit) / Model & \textsc{pow} & \textsc{pow+nsatmos}  & \textsc{pow} & \textsc{nsatmos}  \\
\hline
$N_{\mathrm{H}}$ ($\times10^{21}~\nh$) \dotfill & $6.7 \pm 1.9$ & $7.0 \pm 2.7$ & $2.8^{+4.5}_{-2.2}$ & $4.8 \pm 0.2$ \\
$\Gamma$ \dotfill & $1.7 \pm 0.2$  & $1.7 \pm 0.2$  &  $2.9^{+2.2}_{-1.0}$ &    \\
$kT^{\infty}$ (eV) \dotfill &  & $< 61$   &  & $50 \pm 4$    \\
$\chi^2$/d.o.f. \dotfill& 52.4/58 &  52.3/57 & 10.9/16 &  16.0/17   \\
$F_{\mathrm{X,abs}}$ ($\times10^{-14}~\flux$) \dotfill&  $6.0 \pm 0.5$ &  $6.0 \pm 0.6$ & $0.4 \pm 0.2$ & $0.2 \pm 0.1$   \\
$F_{\mathrm{X,unabs}}$ ($\times10^{-14}~\flux$) \dotfill& $7.8 \pm 0.8$ & $8.0 \pm 1.5$  & $0.6^{+1.8}_{-0.1}$ & $0.9 \pm 0.4$   \\
$L_{\mathrm{X}}$ ($\times10^{32}~[D/7.1~\mathrm{kpc}]^2~\lum$) \dotfill & $4.7 \pm 0.5$ & $4.8 \pm 0.7$ & $0.4^{+1.0}_{-0.1}$ & $0.5 \pm 0.2$  \\
$L_{\mathrm{th}}$ ($\times10^{32}~[D/7.1~\mathrm{kpc}]^2~\lum$) \dotfill&  & $< 3.1$  &  & $1.4 \pm 0.5$  \\
Thermal contribution \dotfill&  &  $<24\%$ &  &   \\
\hline
\end{tabular*}
\tablecomments{Quoted errors represent 90\% confidence levels. $F_{\mathrm{X,abs}}$ and $F_{\mathrm{X,unabs}}$ represent the 0.5--10 keV absorbed and unabsorbed flux, respectively. $L_{\mathrm{X}}$ denotes the 0.5--10 keV luminosity and $L_{\mathrm{th}}$ the 0.01--100 keV thermal luminosity. The bottom row gives the contribution of the thermal \textsc{nsatmos} component to the total unabsorbed 0.5--10 keV model flux. 
}
\end{center}
\end{table*}

\section{Discussion}\label{sec:discussion}

\subsection{The Counterpart to the X-Ray Burst}\label{subsec:counterpart}
\source\ is one of several sources for which \beppo\ detected an X-ray flash without detectable accretion emission \citep[e.g.,][]{cornelisse02}. Follow-up observations were performed with \chan\ for some of these objects (several years later) and revealed weak candidate X-ray counterparts with $L_X \lesssim 5 \times 10^{32}~\lum$. This suggested that they were likely transient neutron star LMXBs that exhibited faint ($L_X \lesssim 10^{36}~\lum$) accretion outbursts when the X-ray bursts were ignited \citep[][]{cornelisse02_chan}. Indeed, several of these sources were later detected during faint accretion outbursts \citep[e.g., \saxeen, \saxtwee, and \saxdrie;][]{hands04,degenaar2008_1828,delsanto2010,altamirano2011_1806}. 

The quality of the WFC data of \source\ was very low, imposing considerable uncertainty on the interpretation of this event \citep[][]{cornelisse02}. However, the similarities between the X-ray flash from \source\ and that of the other sources (i.e., duration, peak flux, overall spectral properties, indication of softening during the decay, and location in the Galactic plane), renders it likely that \source\ too is a bursting neutron star \citep[][]{cornelisse02}. Here we identify two possible quiescent neutron star LMXBs within the WFC error circle, that might provide support for this interpretation.

During our long \xmm\ observation we detected eight weak X-ray sources within the $3.2'$ \beppo/WFC uncertainty of \source. They had 0.5--10 keV luminosities in the range of $L_{\mathrm{X}}\simeq (0.5-5)\times10^{32}~(D/7.1~\mathrm{kpc})^2~\lum$, which is typical for transient neutron star LMXBs in quiescence \citep[see e.g.,][]{jonker2004}. However, based on the X-ray spectral properties and X-ray/optical luminosity ratio's we can discard six of these as potential counterparts to the X-ray burster, leaving only sources 1 and 4 as candidates.

Source 1 was detected at $L_{\mathrm{X}}\simeq 5\times10^{32}~(D/7.1~\mathrm{kpc})^2~\lum$, with a spectrum best described by a power law model with an index of $\Gamma=1.7$. Emission from a neutron star atmosphere with $kT^{\infty}\lesssim61$~eV could contribute up to $\simeq$24\% to the total unabsorbed 0.5--10 keV flux. Only a handful of quiescent neutron star LMXBs have similarly hard X-ray spectra that lack detectable thermal emission. The small LMXB subclass of AMXPs (which display coherent X-ray pulsations during accretion outbursts) exhibit $\Gamma \simeq1.5$ power-law spectra with $\lesssim$40\% attributed to thermal emission \citep[e.g., \sax, \swiftpulsar, and \radiopulsar; ][]{campana2008,heinke2009,degenaar2012_amxp,linares2013_M28}. Their hard spectra are ascribed to their relatively strong magnetic field. The only non-pulsating neutron star with a very hard ($\Gamma \simeq1.7$) quiescent spectrum is the LMXB and X-ray burster \exo\  \citep[e.g.,][]{wijnands2005,degenaar2012_1745}. Its irregular quiescent properties have been interpreted in terms of ongoing low-level accretion. Thus, source 1 could be the counterpart of \source, but it would imply that it is an unusual neutron star (perhaps with a relatively strong magnetic field or exhibiting low-level accretion). 

Source 4, on the other hand, has a soft X-ray spectrum like the majority of quiescent neutron star LMXBs. This source was detected at $L_{\mathrm{X}}\simeq 5\times10^{31}~(D/7.1~\mathrm{kpc})^2~\lum$, and its spectrum can be described by a neutron star atmosphere model with $kT^{\infty}\simeq50$~eV. This relatively low temperature is consistent with the expectations for a neutron star that exhibits faint accretion outbursts \citep[][]{wijnands2012}. We therefore tentatively identify source 4 as the counterpart of the X-ray burst detected with \beppo\ in 1999.

\subsection{The X-Ray Burst Trigger Conditions}\label{subsec:bursttheory}
The duration, repetition rate, and energetics of X-ray bursts depend on the conditions of the ignition layer, such as its thickness, temperature profile, and chemical abundances. These conditions are sensitive to the (local) accretion rate onto the neutron star so that different accretion regimes give rise to X-ray bursts with distinct properties \citep[e.g.,][]{fujimoto81,fushiki1987}. X-ray bursts that ignite in a pure He layer are typically short (seconds), whereas the presence of H in the ignition layer prolongs the duration (minutes).

Upper limits obtained for the accretion emission of the burst-only sources suggests they occupy the lowest accretion regime \citep[$\lesssim0.01~L_{\mathrm{Edd}}$;][]{cornelisse02}. Classical burning theory prescribes that in this regime unstable burning of H should trigger a mixed He/H X-ray burst with a duration of $\simeq100$~s \citep[e.g.,][]{fujimoto81}. This is much longer than the short ($\simeq10$~s) X-ray bursts detected from \source\ and other \beppo\ burst-only sources. This apparent discrepancy was addressed by \citet{peng2007}, who showed that for the low inferred mass-accretion rates sedimentation of heavy elements significantly reduces the amount of H in the ignition layer. This would alter the X-ray burst properties, possibly explaining the observations \citep[][]{peng2007}.

The fact that no persistent X-ray emission was detected above $L_{\mathrm{X}}\simeq8\times10^{32}~(D/7.1~\mathrm{kpc})^2~\lum$ within $\simeq$8 hr after the X-ray burst detection raises the question whether \source\ may have been accreting at $\lesssim 1 \times 10^{-5}~L_{\mathrm{Edd}}$ when the burst ignited.  This is right at the boundary below which theoretical models predict that no X-ray bursts can occur \citep[e.g.,][]{fushiki1987}. We consider this scenario unlikely, because the chance probability of detecting this $\simeq$10-s event would be very low; for a typical ignition column depth of $y\simeq10^{8}~\mathrm{g~cm}^{-2}$ and a mass-accretion rate of $\dot{M}\simeq10^{-13}~\mdot$, the time to accumulate enough material to produce this X-ray burst would be $\simeq$8~yr.  

Regardless whether source 1 or source 4 is the true counterpart, we consider it more likely that \source\ is a transient neutron star LMXB that exhibited an X-ray burst during the decay of a (short) faint accretion outburst. In particular, the behavior of \source\ is strikingly similar to that of the AMXP \swiftpulsar. This source rapidly decayed to a level of $L_{\mathrm{X}}\simeq 1\times10^{33}~\lum$ within a day after it was discovered through the detection of an X-ray burst \citep[][]{wijnands09}. Its behavior remained a puzzle until it exhibited a $\simeq$2-week long accretion outburst with an average 0.5--10 keV luminosity of $L_{\mathrm{X}}\simeq10^{36}~\lum$ several years later \citep[e.g.,][]{altamirano2011_amxp}. This suggests that the peculiar X-ray burst had likely occurred during the decay of its previous accretion outburst. A similar scenario can be envisioned for \sourcefull.

\acknowledgments
ND is supported by NASA through Hubble Postdoctoral Fellowship grant number HST-HF-51287.01-A from the Space Telescope Science Institute. RW acknowledges support from a European Research Council (ERC) starting grant. Support for this work was provided by an \xmm\ GO grant for proposal number 072088. The authors thank the anonymous referee for thoughtful comments that helped improve the clarity of this manuscript.


\bibliographystyle{apj}

\end{document}